\documentclass[useAMS,usenatbib,usegraphicx]{mn2e}
\usepackage{amsmath,graphicx}
\newcommand{\keV}{$\rm\thinspace keV\,$}
\newcommand{\ergscm}{$\rm\thinspace erg\ s^{-1} cm^{-2}\,$}
\makeatletter

\newcommand{\Rmnum}[1]{\expandafter\@slowromancap\romannumeral #1@}
\makeatother

\title[Broad iron L-line and reverberation in 1H0707-495]{Broad iron L-line and X-ray reverberation in 1H0707-495}
\author[A. Zoghbi et. al]{
\parbox{7in}{A. Zoghbi$^{1}$\thanks{E-mail:azoghbi@ast.cam.ac.uk}, A. C. Fabian$^{1}$, P. Uttley$^{2}$, G. Miniutti$^3$, L. C. Gallo$^4$, C. S. Reynolds$^5$, J. M. Miller$^6$ and G. Ponti$^7$}\\
\\
$^{1}$Institute of Astronomy, Madingley Road, Cambridge CB3 0HA\\
$^{2}$School of Physics and Astronomy, University of Southampton, Highfield, Southampton SO17 1BJ\\
$^{3}$LAEX, Centro de Astrobiologia (CSIC--INTA); LAEFF, PO Box 78, E-28691 Villanueva de la C\~{a}nada, Madrid, Spain\\
$^4$Department of Astronomy and Physics, Saint Mary's University, Halifax, NS B3H 3C3, Canada\\
$^5$Department of Astronomy and the Center for Theory and Computation, University of Maryland, College Park, Maryland 20742, USA\\
$^6$Department of Astronomy, University of Michigan, Ann Arbor, Michigan 48109, USA\\
$^7$APC Universit\'{e} Paris 7 Denis Diderot,75205 Paris Cedex 13, France
}
\begin{document}
\maketitle
\label{firstpage}

\begin{abstract}
A detailed analysis of a long \textit{XMM-Newton} observation of the Narrow Line Seyfert 1 galaxy 1H0707-495 is presented, including spectral fitting, spectral variability and timing studies. The two main features in the spectrum are the drop at $\sim7$ \keV and a complex excess below 1 \keV. These are well described by two broad, K and L, iron lines. Alternative models based on absorption, although they may fit the high energy drop, cannot account for the 1 \keV complexity and the spectrum as a whole. Spectral variability shows that the spectrum is composed of at least two components, which are interpreted as a power-law dominating between 1--4 \keV, and a reflection component outside this range. The high count rate at the iron L energies has enabled us to measure a significant soft lag of $\sim 30 $ s between 0.3--1 and 1--4 \keV, meaning that the direct hard emission leads the reflected emissions. We interpret the lag as a reverberation signal originating within a few gravitational radii of the black hole.
\end{abstract}


\section{Introduction}
In a recent paper, \cite{2009Natur.459..540F} reported on the discovery of the first broad iron L line in the Narrow Line Seyfert 1 galaxy 1H0707-495. The interpretation is that the line is broadened by relativistic effects very close to the black hole, where the line is thought to originate.\\
This, along with the already established broad iron K line seen in many AGN, is a further confirmation that the observed X-ray spectra from suppermassive black holes in radiatively efficient AGN are emitted from within few gravitational radii of the black hole (\citealt{1995Natur.375..659T,2007MNRAS.382..194N}, see \citealt{2007ARA&A..45..441M} for a review).

The standard picture is that matter is accreted in the form of an optically thick, geometrically thin disc (\citealt{1973A&A....24..337S}) that radiates mainly in the UV. The usually observed power-law spectrum in X-rays, which is likely to be due to a Comptonising corona and/or the base of jet, is expected to illuminate the disc, producing a characteristic reflection spectrum that is observed alongside the primary power-law (\citealt{1991MNRAS.249..352G}). The main feature in the reflection spectrum is a fluorescent Fe K$\alpha$ line at $\sim 6.4$ \keV, which is expected to be broadened and smeared out if it is emitted close to the black hole. Other features are also expected to be seen in certain circumstances such as high iron abundance (\citealt{2009Natur.459..540F}, this work), while a soft excess is naturally produced for a wide range of ionisation states.

Although alternative interpretations of AGN spectra exist (e.g. \citealt{2009A&ARv..17...47T} for a review), they generally fail to explain some variability properties seen in the data (see the discussion section). Spectral variability and timing analysis provide a completely independent way of studying the physics of emission. That, combined with the spectral fitting, can be used to distinguish between the different models. In particular for the reflection interpretation, the model predicts a time delay, that in principle, is observable between the primary emission and the reflected emission (e.g. \citealt{2001AdSpR..28..267P}). This has been seen for the first time in 1H0707-495 (\citealt{2009Natur.459..540F}) and is discussed further in this paper.

1H0707-495 ($z=0.0411$, hereafter 1H0707), is a highly variable Narrow Line Seyfert 1 (NLS1) galaxy (\citealt{1999ApJS..125..297L}). It has a steep spectrum and shows a strong soft excess, and although these are common among other NLS1, they are extreme in 1H0707 (\citealt{1999MNRAS.309..113V}). The soft excess is traditionally interpreted as disc blackbody emission. However, when large samples of AGN are considered, the measured temperature is higher than the temperature of a standard disc (\citealt{1973A&A....24..337S}), and independent of blackhole mass (\citealt{2004MNRAS.349L...7G}). It appears to be well explained by the blend of relativistically smeared reflection continuum (\citealt{2006MNRAS.365.1067C}; see \citealt{2008MNRAS.386L...1S} for problems with alternative absorption models).

1H0707 gained interest after the discovery of a sharp drop in the spectrum at $\sim 7$ \keV, during the first XMM-Newton observation [O1 hereafter] (\citealt{2002MNRAS.329L...1B}). A second XMM observation (O2) showed a change in the energy of the edge from 7.1 to 7.5 \keV (\citealt{2004MNRAS.353.1064G}). The drop was attributed to either the source being partially covered (\citealt{2002MNRAS.329L...1B,2004PASJ...56L...9T}), or it being the blue wing of a broad iron K$\alpha$ line (\citealt{2004MNRAS.353.1071F}). In 2007, four short (40 ks each) observations were made with XMM (O3 hereafter). A short report on a 4th much longer observation (O4 hereafter) done in 2008 was presented in \cite{2009Natur.459..540F}. In this paper, we present detailed analysis of the results reported there. We also include analysis of the O3 data in some parts of this study which is organised as follows: observation and data reduction are presented in Sec. \ref{observation}, spectral fitting and variability are presented in detail in Sec. \ref{spec_fitting} and \ref{spec_var} respectively. In Sec. \ref{timing}, the lag calculation are shown, with all the results and implications discussed in Sec. \ref{discussion}.
\begin{figure}
\centering
 \includegraphics[height=210pt,angle=270,clip ]{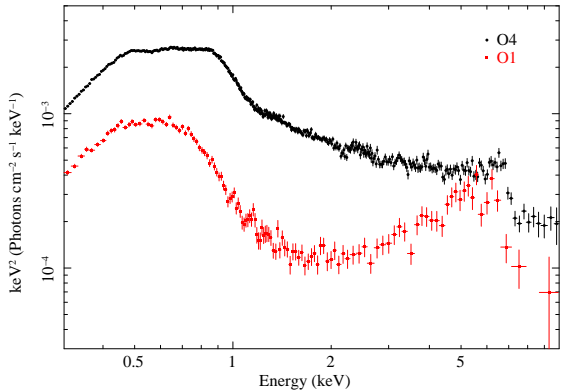}
\caption{Unfolded spectrum of 1H0707 in the 2008 XMM observations (O4) compared to the first observation (O1). The spectra have been unfolded (using \texttt{eeufspec}) against a power-law of index 0. This removes the effect of the detector effective area and shows the main features in the spectrum. O4 represents the spectrum in the `bright' state, and the O1 represents a typical spectrum for the `faint' state.}
\label{fig:spec_all}
\end{figure}

\section{Observations \& Data Reduction}\label{observation}
1H0707-495 was observed for four consecutive orbits using XMM-Newton starting on 2008, January 29th, with a total exposure time of $\sim 500$ ks. The observations were made in the large window imaging mode. The data were analysed using XMM Science Analysis System (\textsc{sas v8.0.0}). The light curve was filtered to remove any background flares. No significant pileup was found in the patterns, which were selected using the condition \textsc{pattern} $\leq 4$ for the spectral analysis, and then relaxed when doing the timing.

Source spectra were extracted from circular regions of radius of 35 arcsec centred on the source, and background spectra from regions on the same chip. The spectra were then grouped to bins with a minimum of 20 counts each. The response matrices were generated using \textsc{rmfgen} and \textsc{arfgen} in \textsc{sas}. For the combined spectrum of the four orbits, the events files were merged together before extracting the spectrum.

The spectra show the presence of two instrumental lines between 8--9 \keV. These appear as two clear emission lines in the background spectra, and as absorption lines in the background-subtracted source spectra. Their energies are 8.0 and 8.9 \keV respectively and can be identified as Cu-K lines originating from the electronics circuit board under the CCD (\citealt{2004A&A...414..767K}). We tried different background regions to avoid the circuit board, but we found that for the best signal to noise ratio, it is optimal to get a large background region and include two absorption gaussians at 8.05 and 8.91 \keV in all fits. All results reported have been checked with spectra with cleaner background, but higher noise.

Spectral fitting was performed using {\sc xspec v12.5.0} (\citealt{1996ASPC..101...17A}). All quoted errors on the model parameters correspond to a 90 per cent confidence level for one interesting parameter (i.e. a $\bigtriangleup\chi^2 = 2.71$), and energies are given in the rest frame of the source (unless stated otherwise). The quoted abundances refer to solar abundances in \cite{1989GeCoA..53..197A}.
\begin{figure}
\centering
 \includegraphics[height=240pt,angle=270,clip ]{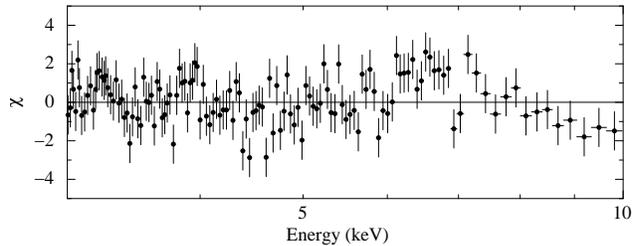}
\caption{A plot of residuals in terms of sigma for the simple power-law + edge model fitted to the high energy band.}
\label{fig:pc_simple_edge}
\end{figure}

\section{Spectral Analysis}\label{spec_fitting}
Fig. \ref{fig:spec_all} shows the spectra from the O4 observation along with an earlier observation (O1) for comparison. O4 shows a typical spectrum when the source is in a `bright' state, which is similar to O2, and three of the four O3 (not shown). The O1 spectrum is typical when the source is faint which is similar to the first segment of O3. In what follows, the spectrum is explored in detail looking initially at hard energies ($>3$ \keV), before including softer energies and looking at the whole spectrum.
\subsection{Hard Spectrum}\label{hard_fit}
The main feature in the hard spectrum of 1H0707 is the sharp drop around 7 \keV. To characterise the feature we fitted the 3-10 \keV spectrum with a power-law and an edge. The fit was resonable ($\chi^2 = 313$ for 258 degrees of freedom [d.o.f]). The edge energy is $E = 7.31^{+0.12}_{-0.06}$ \keV with a depth of $\tau = 0.89^{+0.12}_{-0.13}$.
The energy of the drop changed significantly between earlier XMM observations, from $7.1\pm0.1$ \keV to $7.5\pm0.1$ \keV for the O1 and O2 observations respectively (\citealt{2004MNRAS.353.1064G}). The energy in O4 is between the two, and is more consistent with the O2 observation. The depth is also consistent within errors with that reported for O2 ($\tau = 0.84^{+0.25}_{-0.22}$).
\begin{figure}
\centering
 \includegraphics[height=240pt,angle=270,clip ]{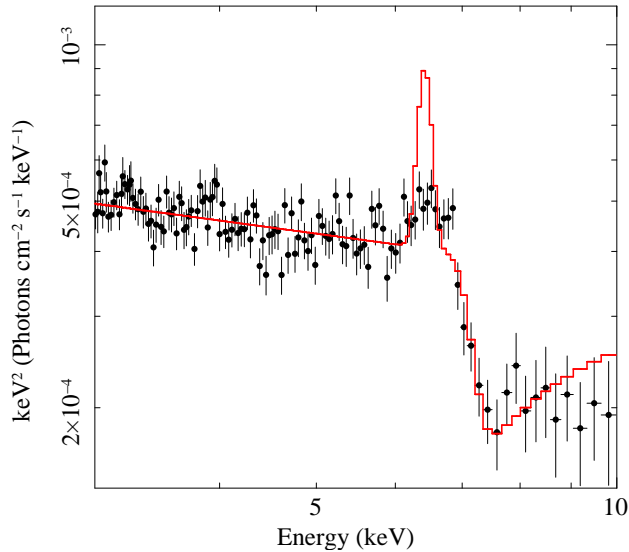}
\caption{The spectrum of 1H0707 in the hard band (data points). The solid line is the model (a neutral edge from an outflow with $v=0.03c$) fitted to the data plus an emission line expected from simple physics given the edge depth. If the edge is due to ionised material, the fluoresence yield increases and the line should be stronger.}
\label{fig:pc_missing_flux}
\end{figure}

Although the general shape of the drop is matched, few residuals clearly remain (see Fig. \ref{fig:pc_simple_edge}), in particular the positive excess at $\sim 6.5$ \keV and the apparent absorption at $\sim 7$ \keV. The former is possibly due to iron K$\alpha$ emission. The latter might be caused by H-like iron absorption. The fit improves slightly if a smoothed edge is used instead, ($\Delta\chi^2=11$ for 1 less d.o.f), giving a width of $171^{+10}_{-7}$ eV, consistent with O1 and O2.
If the spectral drop is interpreted as an iron K-edge, a corresponding emission line is expected. Adding a narrow gaussian line improves the fit slightly, it has energy of $6.73^{+0.09}_{-0.10}$ \keV. The fit however, still leaves residuals below $\sim 6.7$ \keV, which indicates that the line is broader. If $\sigma$ is allowed to vary, it has a value of $\sigma=295^{+211}_{-86}$ eV. 
For ionisation states where resonant absorption is not important, a photon absorbed in the K-edge is followed by the emission of a fluorescent line with a probability corresponding to the fluorescent yield. Its value is 0.34 for neutral iron and an absorber that covers $4\pi$ solid angle as seen by the emitter. The observed line energy of 6.73 \keV is inconsistent with it originating in the same absorbing material ($E_{edge}=7.3$ \keV), regardless of ionisation or blue/redshift considerations. The $90$ per cent upper limit of the flux from the emission line in 1H0707 is $7.66\times10^{-15}$ \ergscm, 
while the absorbed flux is $8.54\times10^{-14}$ \ergscm (the edge and line energies are now $\sim$ 7.4 and 6.7 \keV respectively). This implies that $<9$ per cent of the absorbed flux is re-emitted as florescence, while $\sim30$ per cent is expected (\citealt{1987ApJ...320L...5K}). Fig. \ref{fig:pc_missing_flux} shows the line expected based on the absorbed flux if $4\pi$ covering fraction is assumed. Although the ratio of expected to observed line flux might be reduced if the covering fraction is less ($\sim 20$ per cent as implied by the data), it should be noted that this is just a lower limit given the ionisation assumed for the absorber.
\begin{figure}
\centering
 \includegraphics[width=220pt,clip ]{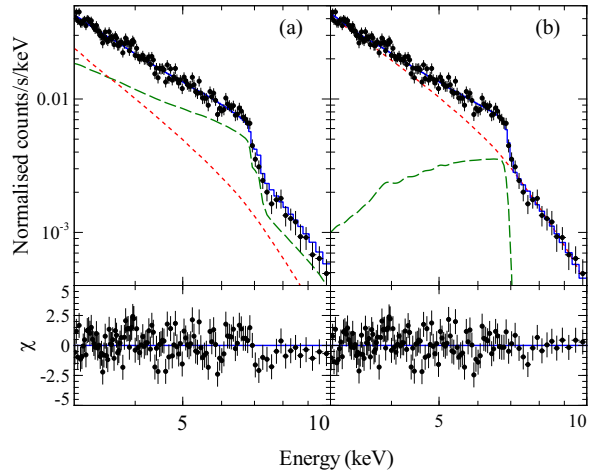}
\caption{A plot of residuals in terms of sigma for the two models considered for the hard spectrum. (a): a partially covered power-law (green long-dashed) by an ionised absorber, and (b): a broad K$\alpha$ line. (green long-dashed). For both plots, the dashed (red) line is power-law and the continuous (blue) line is the total. The data have been rebinned for display.}
\label{fig:pc_ref_hard}
\end{figure}

If iron abundance is included in the fit as a free parameter (using \texttt{zvfeabs} in \textsc{xspec}), a lower limit ($90$ per cent confidence) on the abundance of $\sim 18\times$ solar is found, while the best fit value is higher, similar to earlier observations (e.g. \citealt{2002MNRAS.329L...1B}). The best fitting power-law index is $\Gamma = 3.2\pm0.1$.
\cite{2004PASJ...56L...9T} used a broken power-law instead. That reduces the abundance slightly, but the fit indicates that the steepness introduced by the broken power-law is mainly driven by the deep drop, and not by an actual break in the power-law below 7 \keV. A broken power-law fitted to energies below the drop does not fit better in a statistical sense than a simple power-law.

To gain more physical insight into the K-shell absorber, we used the XSTAR photoionisation code (\citealt{1996ApJ...465..994K}) to generate a grid of models to fit the data. We fixed the abundances at solar values except for iron which was left free, the illuminating flux was a steep power-law of index 3. Other fitting parameters include column density ($N_{\rm{H}}$), ionization parameter ($\xi=4\pi F/n$ where $F$ is the flux at an absorber of density $n$) and covering fraction. If the iron abundance is fixed at solar, the fit is not good ($\chi^2_\nu=1.7$ for 258 d.o.f) and improves if it is allowed to vary, it has a lower limit of $\sim 14\times$ solar (panel (a) in Fig. \ref{fig:pc_ref_hard}). At solar values, the high column density required to fit the drop gives too much continuum curvature below the edge.
The best fit absorption model (which does not include any re-emitted components) has $N_{\rm{H}}\sim 6\times10^{22}\rm{cm}^{-2}$, $\rm{log}\xi\sim3.0$ and a partial covering fraction of $\sim 0.7$. This means that the absorber covers 20 per cent of the sky as seen by the source, 70 per cent of it is in our line of sight. Given the high ionisation parameter, the model also predicts a K$\beta$ UTA (e.g. \citealt{2002ApJ...577L.119P}), which is not seen in the data. If the absorber is allowed to be moving, an outflow velocity of $\sim0.01c$ is found ($\Delta\chi^2=12$ for 1 extra d.o.f).
\begin{figure}
\centering
  \includegraphics[width=290pt,angle=270,clip ]{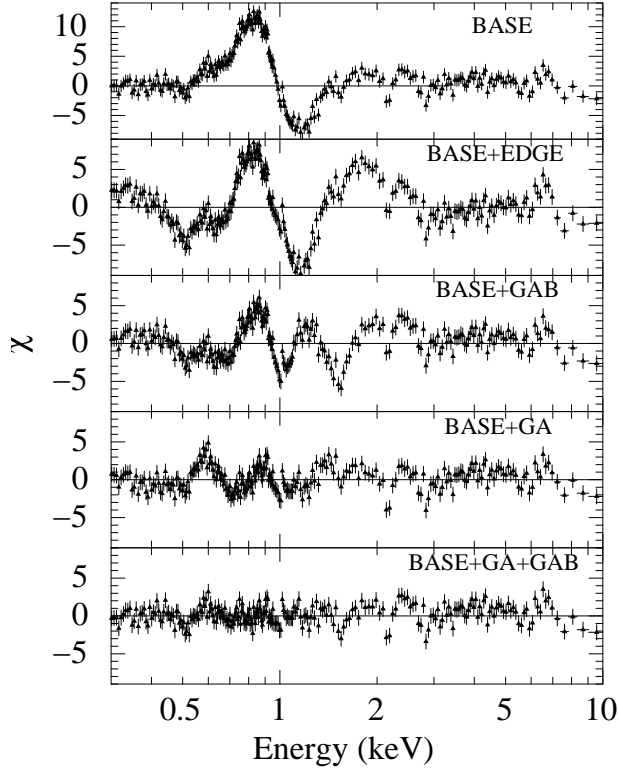}
\caption{A plot of residuals in terms of sigma for a model consisting of a partially covered power-law by a photoionised gas + disc blackbody (BASE), plus different phenomenological models, EDGE: Absorption edge, GAB: absorption gaussian, GA: emission gaussian.}
\label{fig:base_plus_diff}
\end{figure}

Alternatively, the spectral drop can be attributed to emission rather than absorption. The prominent emission feature in these energies would be that of iron K$\alpha$. The line has to be very broad to explain the drop and the excess emission below $\sim 6$\keV ($\sigma\sim0.97$ \keV, $E\sim5.9$\keV and equivalent width of 1.2 \keV if fitted with a simple gaussian). 
If a relativistically broadened line is used (\texttt{laor} of \citealt{1991ApJ...376...90L} in \textsc{xspec}), a good fit is found ($\chi^2_\nu=1.03$ for 255 d.o.f, see Fig. \ref{fig:pc_ref_hard}). The energy, inclination and inner radius of the line were found to be $E=6.19^{+0.07}_{-0.04}$\keV (rest frame), $i=53^{+2}_{-9}$ degrees and $r_{\rm in}=1.56^{+0.08}_{-0.09} r_{\rm g}$ respectively, where $r_{\rm g} = GM/c^2$ is the gravitational radius of the black hole. An equally good fit is found if the energy is fixed at 6.4 \keV with very small changes to the other parameters. The power-law index was $\Gamma=2.9\pm0.1$. No other significant emission or absorption feature is present in the spectrum (Fig. \ref{fig:pc_ref_hard}).

To see what effect the continuum would have on the broad line fit parameters, and to account for other emission expected with iron line, we used the self-consistent reflection model of \cite{2005MNRAS.358..211R} \textsc{reflionx}. This was convolved with the \texttt{kdblur} kernel (in \textsc{xspec}) to account for relativistic effects. The model has as free parameters the iron abundance, ionization $\xi$ and the illuminating power-law index $\Gamma$ (which is linked to the main power-law index in the fit). \texttt{kdblur} parameters include $r_{\rm in}$, $i$ and $q$ (where emissivity $\propto r^{-q}$), in addition to $r_{\rm out}$ which was fixed at 400 $r_{\rm g}$. The fit was very similar to that of the \textit{laor} line. It has $r_{\rm in} = 1.4\pm0.1 r_{\rm g}$, $q=8.9^{+1.1}_{}$ and $i = 45\pm1$ degrees. The iron abundance was found to have a lower limit ($90$ per cent) of $ 7\times\rm{solar}$, similar to previous work (\citealt{2004MNRAS.353.1071F}). This indicates that most of the emission is coming from a small region very close the black hole.

\subsection{Soft Spectrum}\label{soft_fit}
The spectrum of 1H0707 does not show the strong absorption below $\sim 3$\keV which would be expected given the absorbing column density inferred from the deep Fe K$\alpha$ edge. This means that, in the absorption interpretation, the absorber only partially covers the source.
The best fitting model consisting of a partially covered power-law was extended down to 0.3 \keV. A strong excess at soft energies ($< 1$\keV) is present. This feature is common in other NLS1 (e.g \citealt{2004MNRAS.349L...7G}). The standard picture is to attribute this to disc emission. However, the observed disc temperature ($\sim 150$\keV) is higher than expected from a standard thin disc, and is independent of the black hole mass (\citealt{2004MNRAS.349L...7G, 2006MNRAS.365.1067C}). Including a multi-colour disc blackbody improves the fit slightly (with temperature $T=0.14$ \keV), however it leaves strong residuals at $\sim 1$\keV (Fig. \ref{fig:base_plus_diff} top, this model is referred to hereafter as BASE). The same residuals were found by \cite{2004MNRAS.353.1064G} in this source, and by adding two gaussians (an emission and an absorption), the fit improves significantly ($\Delta\chi^2\sim 7500$ for 6 extra d.o.f, Fig. \ref{fig:base_plus_diff}). \cite{2004MNRAS.353.1064G} interpret the absorption as being due to a blend of resonance lines, mostly from ionized L-shell iron, while the emission arises from an extended warm medium outside the line of sight. This seem rather arbitrary and, as suggested by the authors, a better explanation is required.
\begin{figure}
\centering
  \includegraphics[width=160pt,angle=270,clip ]{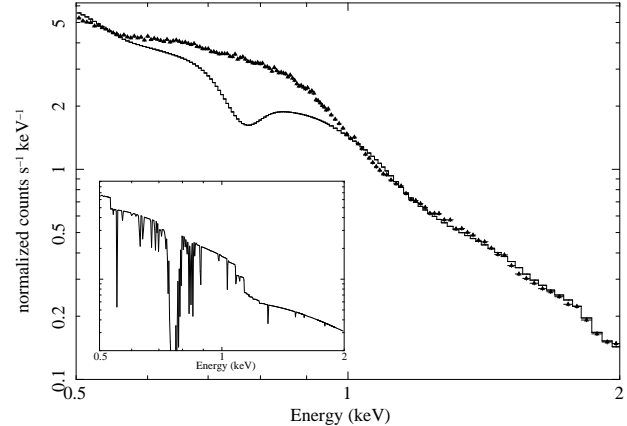}
\caption{The photoionisation absorption model fitted to the data. Main plot: the data and the model that under-predicts the emission at $\sim 0.8$ \keV. Inset: a detailed plot of the model showing the UTA features due to iron.}
\label{fig:show_uta}
\end{figure}

The residuals are clearly not caused by absorption from {\textsc OVII \rm{or} OVIII}, and including edges fixed at 0.74 and 0.87 \keV does not improve the fit at all. If the edge energy is allowed to vary it shifts to higher values but does not improve the fit by much (BASE+EDGE in Fig. \ref{fig:base_plus_diff}). The edge has energy $E=1.011\pm0.004$ \keV and $\tau=2.7\pm0.4$. If this is a blue-shifted {\textsc{OVIII}} edge, the emitting material would have to be moving at very high velocities ($\sim 0.16c$). A better fit is found if an absorption line is used instead ($\Delta\chi^2=2181$ for the same number of d.o.f, with $E=1.163\pm0.004$ and $\sigma=0.0144\pm0.005$). However, clear systematic residuals are left which indicates that the model is not consistent with the data (BASE+GAB in Fig. \ref{fig:base_plus_diff}).

If the feature is modelled as an emission rather than absorption, a much better fit is found ($\chi^2=1495$ for 1104 d.o.f, BASE+GA in Fig. \ref{fig:base_plus_diff}). A fit with a gaussian emission line gives a line energy of $E=0.799\pm0.003$\keV and a width of $\sigma=127\pm6$ eV. This is most likely due to iron L-shell lines.

The super-solar abundances implied from the fits to the hard spectrum, can (or is expected to) have other possible effects on the spectrum. In particular, if the drop at $\sim 7$ \keV is caused by absorption, then one might expect to see a similar drop caused by the L-shell edge ($E_{\rm{edge}}>0.85$ \keV). A simple phenomenological model consisting of a power-law, blackbody and two edges reproduces the general shape of the spectrum. This model is different from BASE+EDGE shown in Fig. \ref{fig:base_plus_diff}), where the full spectrum generated by XSTAR is used. This simple phenomenological model is unphysical, because it ignores the effect of the continuum, which needs to be modelled properly in any absorption scenario. 
\begin{figure}
\centering
  \includegraphics[width=160pt,angle=270,clip ]{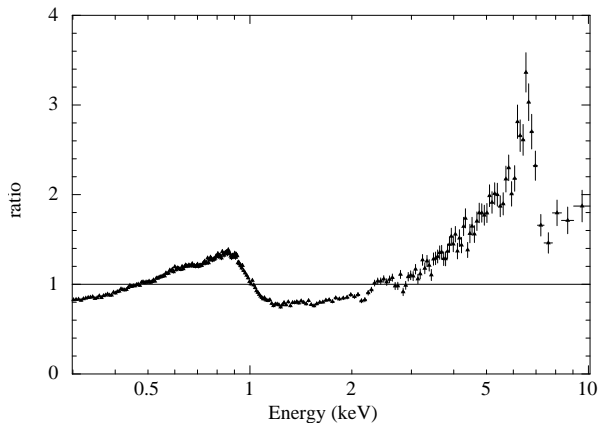}
\caption{A ratio plot of the data to a simple absorbed power-law. Galactic absorption was fixed at $5\times10^{20}\rm{cm}^{-2}$.}
\label{fig:simple_pow_rat}
\end{figure}

The fact that an absorber producing the K-edge has to be partially covering the source imply that the two features at 7 and 1 \keV \textit{cannot} be due the same absorbing cloud. The best fitting (partially covered) XSTAR model of Sec. \ref{hard_fit} was extended to lower energies, a second ionised absorber was added to try to account for the $\sim 1$ \keV feature. A blackbody component was also included to account for the soft excess. The  parameters of the second absorber were allowed to vary. No good fit was found. The main reason is that the model tries to fit the apparent drop below 1 \keV with an L-shell iron edge, requiring high abundances, however, as Fig. \ref{fig:show_uta} shows, this predicts a drop between 0.7 and 0.9 \keV due to the M-shell unresolved transition array (UTA) (\citealt{2009Natur.459..540F}). This is a blend of numerous absorption lines arising from the photoexcitation of ions Fe\Rmnum 1 -- Fe\Rmnum {16} mainly produced by $2p-3d$ transitions (\citealt{2001ApJ...563..497B}). It should be noted that O\Rmnum7 and O\Rmnum8 edges are also predicted by the model but not seen in the data, particularly for smaller values of the iron abundances. \cite{1999ApJ...517..108N} showed that similar features in an ASCA observation of IRAS 13224-3809 can be reproduced by a warm absorber that is illuminated by a steep power-law. This, however, predicts narrow absorption lines which are not present in the RGS data of 1H0707 (see Fig. 3 in \citealt{2009Natur.459..540F}).

The conclusion of this analysis is that the spectrum around 1 \keV can \textit{not} be described by absorption. \cite{2009Natur.459..540F} has pointed out that this spectrum can be due to a \textit{broad} Fe-L line. This is expected if the high energy drop is interpreted as the blue wing of a broad K$\alpha$ line. Their ratio plot (their Fig. 9, which is reproduced here in Fig. \ref{fig:simple_pow_rat}) of the data to a simple model consisting of a power-law with galactic absorption show clearly two main broad line-like residuals. If, as a phenomenological model, two \textit{Laor} lines are added to the fit, a good description is found (though not statistically acceptable). When the fit parameters of the two \textit{Laor} lines are linked, the fit parameters are: $r_{\rm{in}}=1.82\pm0.04\ r_{\rm g}$ and $Incl. = (51^{+0.6}_{-1.7})^{\circ}$\footnote{The fit parameters are slightly different from those reported in \cite{2009Natur.459..540F} because only the first orbit of the data was used in getting their parameters, whereas all four orbits are used in this work.}. The emissivity index is $5.3\pm0.1$. The energies of the lines are $\sim0.91$ and $\sim 6.0$ \keV (rest-frame) for the L and K lines respectively. These lines are not exactly at the expected energies because the continuum is not modelled properly, and this emphasises the point that these lines cannot be fitted separately, and need to be modelled as part of the whole reflection continuum. If the line parameters between the two lines are allowed to vary independently, they still give similar and consistent results (K-line: $r_{\rm{in}}=1.4^{+0.1}_{-0.2}\ r_{\rm g}$, $Incl. = 57 \pm5^{\circ}$, L-line: $r_{\rm{in}}=1.41\pm0.05\ r_{\rm g}$, $Incl. = 55.6 \pm1.0^{\circ}$, the emissivity index is $5.8\pm0.2$ in both cases).
\begin{figure}
\centering
\includegraphics[width=155pt,angle=270,clip ]{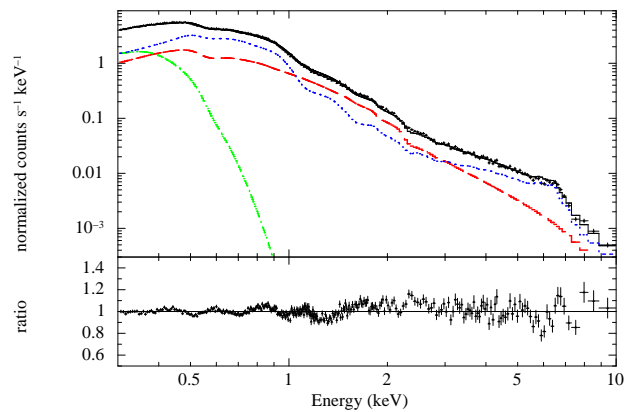}
\caption{The best fitting reflection model with data to model ratio. Long-dashed (red) line is the power-law, dot-dashed (green) is the blackbody and dotted (blue) line is the blurred reflection component, the black line is the total. The reflection fraction (ratio of the reflection to power-law fluxes) is $1.3$.}
\label{fig:emission_ref_all}
\end{figure}

To account for the whole reflection spectrum expected to accompany the two broad lines, we again used the relativistically-blurred self-consistent reflection model \textsc{reflionx} (\citealt{2005MNRAS.358..211R}). The best fitting model is shown in Fig. \ref{fig:emission_ref_all} (this is similar to Fig. 10 in \citealt{2009Natur.459..540F} shown there for the first orbit only). The model gives a very good description of the data. A low temperature black body was required to fit a small excess below $\sim 0.5$ \keV. This is different from the blackbody generally used which have higher temperature ($\sim 0.15$ \keV, \citealt{2004MNRAS.349L...7G}). The temperature of the disc blackbody in our fits is $kT \sim 50$ eV, which is more consistent with what is expected from a thin accretion disc (\citealt{1973A&A....24..337S}).
\begin{figure*}
\centering
\includegraphics[width=440pt,clip ]{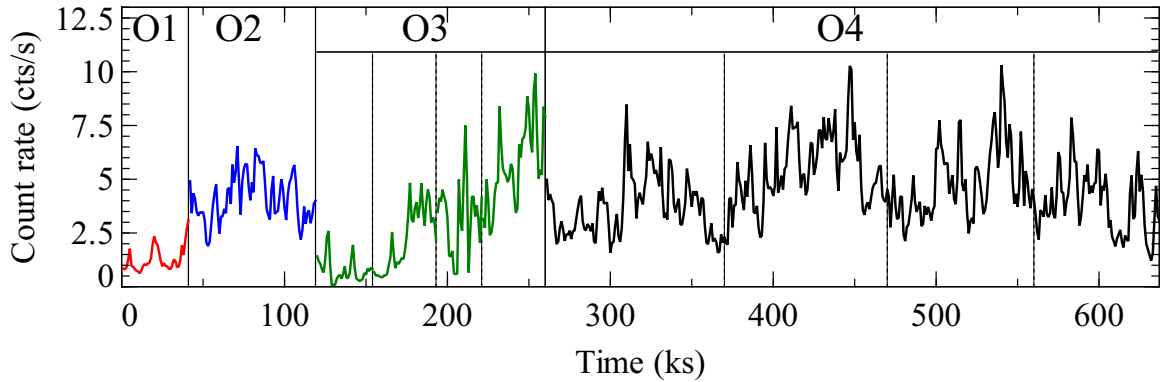}
\caption{The light curve of the current observation compared to previous XMM observations. O1, O2 and O3 refer to the first, second and third observations respectively. O4 is the current observation. O1 was made in full frame mode, while O2, O3 and O4 were made in large window mode. The vertical dashed lines separate the different segments of the observation.}
\label{fig:lc_all}
\end{figure*}

The fit parameters for the blurred reflection model were consistent with the two-\textit{laor} fit. The inner radius was $r_{\rm{in}}=1.23^{+0.07}_{-0.0}\ r_{\rm g}$, with a disc inclination of $Incl. = 58.5^{+0.8}_{-0.7}\ ^{\circ}$. The emissivity index was $6.6^{+1.9}_{-1.1}$. The combination of small inner radius and emissivity index is an indication that most of the emission originates very close to the black hole, and that explains the very broad iron lines. The power-law has a steep spectrum of $\sim 3.2$, this is in line with the established fact NLS1 have steeper than usual X-ray spectra (\citealt{1997MNRAS.285L..25B}). The residuals in the fit between 2 and 5 \keV can be explained by ionisation changes of the spectrum. The data fitted is the time average of four orbits, and trying to parametrise a highly variable spectrum with a single ionization model might not be physical. If an extra reflection component is added, with all parameters linked to the first except for the ionisation (and normalisation), the excess disappears. For the small residuals at $\sim 1$ \keV, most of them are due to ionisation changes, in addition to elements other than iron having non-solar values. The process responsible for producing the high iron abundances, inferred from the K and L lines, is also expected to affect the abundances of other elements.

\section{Spectral Variability}\label{spec_var}
The source has shown significant variability in the past. In particular, the first observation of 2000 (\citealt{2002MNRAS.329L...1B}) seems to have caught the source in a low state compared to the second observation (\citealt{2004MNRAS.353.1064G}). The source in the current observation is in a similar flux state to that of O2, which is clear from the fit parameters above, and the light curve shown in Fig \ref{fig:lc_all}.
\begin{figure}
\centering
\includegraphics[width=240pt,clip ]{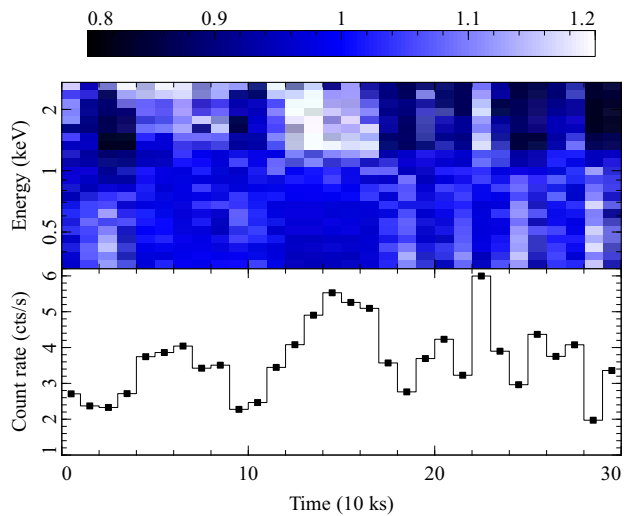}
\caption{Time-resolved spectra for 10ks segments. Top: a 2D image of the ratio of the spectrum in each segment to the average spectrum, as a function of energy and time, a typical error in these points is $\pm0.03$. Bottom: the corresponding count rate in each segment (0.3-10 \keV).}
\label{fig:time_resolved_image}
\end{figure}

\subsection{Time-resolved spectra}\label{time_resolved_spectra}
The spectral fitting presented in Sec. \ref{spec_fitting} showed that a reflection spectrum originating close to the black hole can fit the data very well, with alternative absorption models failing to account for the complexities at $\sim 1$ \keV. In this section we investigate the variability of the spectrum with time. This can provide more constraints on the spectral models.

The O4 data have been divided into segments of 10-ks length. A ratio of each spectrum to the average was calculated. This will highlight the variability pattern in the spectrum on the time-scale probed. This was achieved by fitting the time-averaged spectrum with a multi-spline model to get the best description of the data, and then fitting it to the segments allowing only a multiplicative constant to vary. The spectra in each segment were binned using \textsc{grppha} so that each bin has a minimum of 20 counts per bin (with similar bins between segments), then each ten energy bins were grouped together so the errors are further reduced.
\begin{figure}
\centering
\includegraphics[width=180pt,clip ]{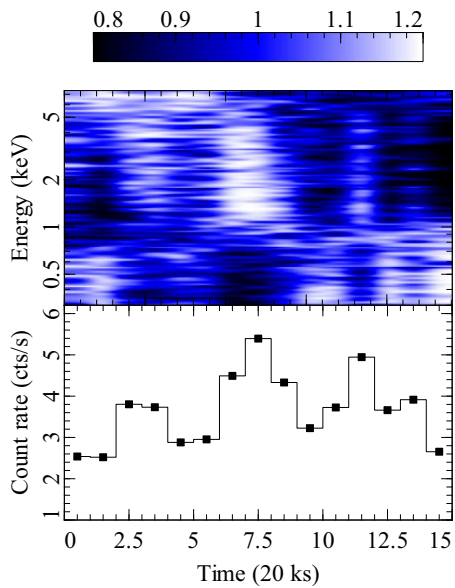}
\caption{Time-resolved spectra for 20ks segments. The data is presented in a similar way to that in Fig. \ref{fig:time_resolved_image}. The data image have been slightly smoothed to show the patterns more clearly.}
\label{fig:time_resolved_image_20}
\end{figure}

The results are shown in Fig. \ref{fig:time_resolved_image}, alongside the corresponding count rate from each segment for $E<3.5$ \keV where the noise effect of noise is minimum. The first point to notice in the variability is the apparent line dividing the spectrum at $\sim 1$ \keV. The variability patterns below and above 1 \keV appear anti-correlated. This may be an indication that there is spectral pivoting at 1 \keV.

The other noticeable pattern is the fact that residuals above 1 \keV match the total count rate very well (Fig. \ref{fig:time_resolved_image}). As the total flux gets higher, the residuals above 1 \keV increase and those below get smaller. The opposite is also true.

To investigate the possibility of spectral pivoting at 1 \keV, we extended the analysis to include higher energy bins. This was not possible with 10 ks segments, so we used 20 ks instead. Fig. \ref{fig:time_resolved_image_20} shows the results. The energy now extends to $\sim 8$ \keV. The data were binned in a similar way to Fig. \ref{fig:time_resolved_image} except for grouping every 2 instead of 10 energy bins.

The figure shows similar patterns to those in Fig. \ref{fig:time_resolved_image}, but now the variability pattern above 1 \keV does not continue up to higher energies, and there appear to be another cut at $\sim 5$ \keV. Also, there seem to be indications that variability above $\sim 5$ \keV is similar to that below 1 \keV.

The fact that the variability pattern does not extend above 5 \keV is inconsistent with spectral pivoting. It can rather be explained if the spectrum is composed of at least two components, one dominating between 1 and 5 \keV (hereafter C$_{<1-5>}$), which is highly correlated with the total flux of the source. The other component dominates outside the $\sim 1-5$ \keV range (hereafter C$_{>1-5<}$).

This model-independent description of the variability is fully consistent with the reflection model presented in Fig. \ref{fig:emission_ref_all}. A power-law component (PLC) dominates between $1-5$ \keV and a reflection-dominated component (RDC) is present outside this range. This picture is similar to that reported for MCG-6-30-15 (\citealt{2003MNRAS.340L..28F, 2004MNRAS.348.1415V})
\begin{figure}
\centering
\includegraphics[width=240pt,clip ]{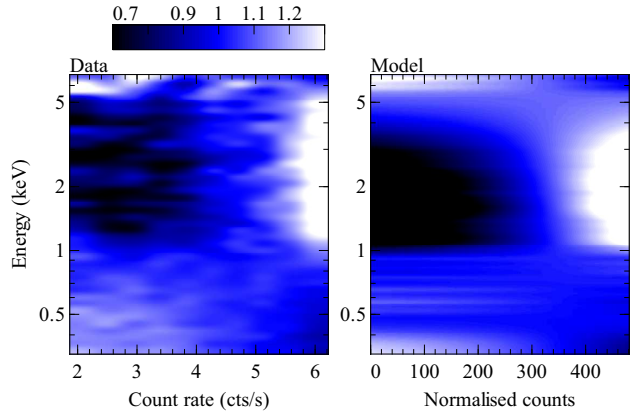}
\caption{\textit{Left}: Flux-resolved spectra for 10 flux segments, each column in the image is a ratio of the segment spectrum to the average. The data image have been slightly smoothed to show the patterns more clearly. The error in points is of order 0.03. \textit{Right}: The same as in the left panel but obtained from simulations in which the power-law varies up and down by 20 per cent keeping the reflection constant (see. Fig. \ref{fig:emission_ref_all}).}
\label{fig:flux_resolved_image}
\end{figure}

\subsection{Flux-resolved spectra}\label{flux_resolved_spectra}
Another way to explore the variability is through flux-resolved spectra (e.g. \citealt{2004MNRAS.348.1415V}). We carried out similar analysis to that in Sec. \ref{time_resolved_spectra}, but now the spectra are extracted from flux segments instead of time segments. The light curve was divided into ten flux bins with each having equal number of points. The spectra from each segments was then compared to the average spectrum.\\
Fig. \ref{fig:flux_resolved_image} (left) shows the result. At low fluxes the spectrum shows an excess (compared to the average) below 1 keV and above 5 \keV. As the flux increases the excess is shifted, and most of it is between 1 and 5 \keV. This, as was pointed out in Sec. \ref{time_resolved_spectra} (see also \citealt{2003MNRAS.340L..28F, 2004MNRAS.348.1415V}), can be easily explained if the spectrum if composed of two components (C$_{<1-5>}$ and C$_{>1-5<}$), and seem to be consistent with a power-law component (PLC) between 1 and 5 \keV and a reflection-dominated outside the range, as described in Sec. \ref{spec_fitting}.

The two-component interpretation of the spectral variability has been suggested by different authors for Seyfert galaxies (\citealt{2003MNRAS.340L..28F,2003MNRAS.342L..31T,2004MNRAS.348.1415V}). 1H0707 seems to show similar patterns. This simple picture is also consistent with the best fitting model shown in Fig. \ref{fig:emission_ref_all}. For example, Fig. \ref{fig:flux_resolved_image} (right) shows the result of applying the same analysis procedure of Sec. \ref{flux_resolved_spectra} to a simulated set of spectra, produced by taking the best fitting model of Fig. \ref{fig:emission_ref_all}, and generating different spectra by changing the normalisation of the power-law randomly by 20 per cent, and it shows clearly that the patterns are reproduced very well.

\begin{figure}
\centering
\includegraphics[width=210pt,clip ]{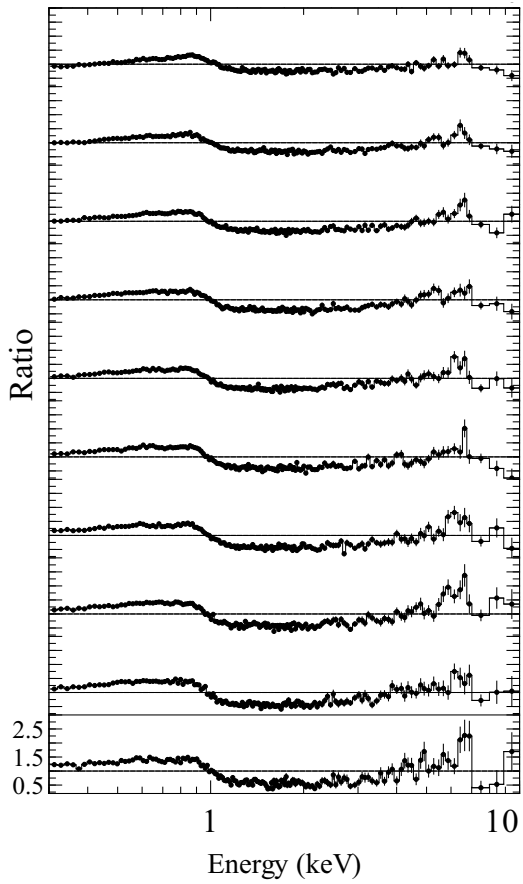}
\caption{Ratio of the spectra in each flux segment to an absorbed power-law of index 3. Flux increases up. The y scale is shown for the first plot, others are plotted on the same scale. The horizontal line with each plot represents the ratio of 1.}
\label{fig:flux_vertical_plot}
\end{figure}

Another representation of the spectral changes as a function of flux is shown in Fig. \ref{fig:flux_vertical_plot}. It shows a ratio of the spectrum in each flux segment to an absorbed power-law of index 3.

What is apparent is that the spectral drops, both at 1 and 7 \keV become weaker as the flux increases. This, combined with the earlier conclusion that the spectrum below 1 \keV and above 5 \keV represents the same component, point to one conclusion: C$_{<1-5>}$ is well correlated with the total flux, while C$_{>1-5<}$ is more constant, the contrast between the two components causes the observed effect.
\begin{figure}
\centering
\includegraphics[height=240pt,angle=270,clip ]{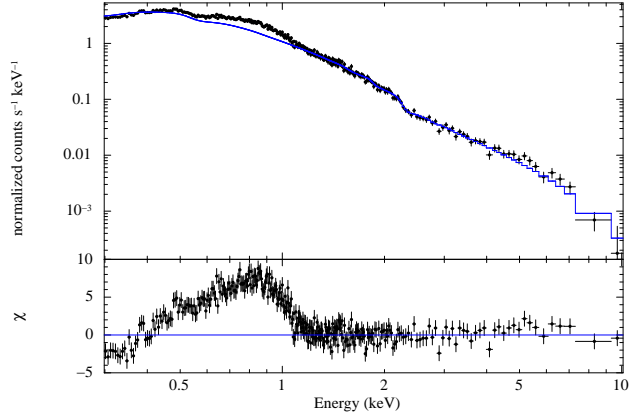}
\caption{The difference spectrum between high and low states, defined as that above and below the mean respectively, along with and absorbed power-law fit and the residuals. This represent the variable component.}
\label{fig:diff_spec}
\end{figure}

\subsection{Difference Spectrum \& the Nature of the Spectral Components}
We showed earlier, using flux and time-resolved spectra, that the spectrum appears to be composed of two components. To find their nature, we plot the difference spectrum in Fig. \ref{fig:diff_spec} (similar to Fig. 7 in \citealt{2009Natur.459..540F}). This is produced by taking the difference between two flux segments (above and below the mean count rate). This way, if the spectrum is composed of two components, a constant and a variable, the constant part is subtracted and only the variable component is left (\citealt{2003MNRAS.340L..28F}).

The variable component cannot be fitted with a single power-law ($\chi^2_\nu=2.9$), however if the power-law is fitted only above 1.5 \keV, a very good fit is found with an index of 3.2 ($\chi^2=218$ for 270 d.o.f, see Fig. \ref{fig:diff_spec}). 

Although interpreting the physical origin of the variable component might not be straight forward, the constant component is consistent with the C$_{>1-5<}$ discussed earlier, and can be easily identified as a reflection dominated component (RDC), characterised by the two broad K and L lines and the accompanying reflection continuum.

The variable component is well-fitted with a power-law above $\sim 1.5$ \keV, and is similar to that found in other AGN (MCG-6-30-15 \citealt{2004MNRAS.348.1415V}, Mrk 766 \citealt{2007A&A...463..131M}, although both sources show signatures of warm absorption, unlike the case here). We interpret the variability below 1 \keV as being due ionisation changes (\citealt{2009Natur.459..540F}). The spectrum, as the best fit shows, is dominated at these energies by reflection, mainly iron L emission and emission from other elements, and as the power-law changes, we would expect the reflector's ionisation to change too.
\begin{figure}
\centering
\includegraphics[width=240pt,clip ]{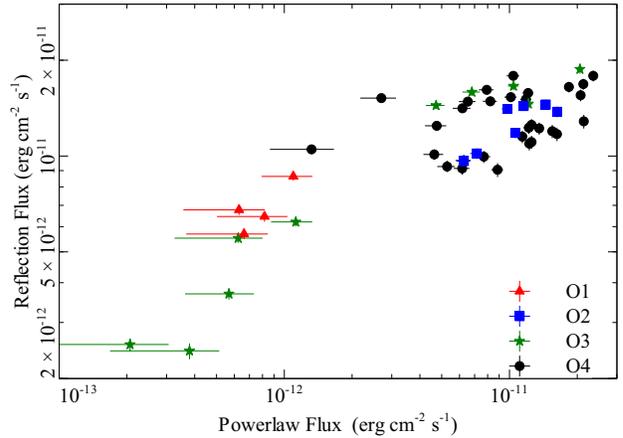}
\caption{A plot of the reflection flux vs power-law flux (both in 0.3--10 \keV) changes in the 10 ks segments. The best fitting model in Fig. \ref{fig:emission_ref_all} was fitted to each segment, allowing the normalisations, photon index and ionisation to change.}
\label{fig:pow_ref_all}
\end{figure}

To investigate this interpretation further, we fitted each of the time-resolved spectra discussed in Sec. \ref{time_resolved_spectra} individually with the best fitting model shown in Fig. \ref{fig:emission_ref_all}. The normalisations of the power-law and reflection components were allowed to vary, as well as the power-law index and the ionisation of the reflection. Other parameters are not expected to vary on the time-scales probed here so they were fixed. To increase the ranges of parameters, we also included data from the two earlier XMM observations of the source (O1 and O2).
Fig. \ref{fig:pow_ref_all} shows a plot of the flux changes of the power-law and reflection components inferred from the fits to 10-ks segments. The figure shows that the two components are correlated, particularly if the O1 low state data (red triangles) are also included. It is also apparent that despite this correlation, most of the variability is driven by the power-law component. Two orders of magnitude change in the power-law corresponds to a factor of $\sim 3$ change in the reflection component. This is generally similar to the pattern seen in MCG-6-30-15 (\citealt{2004MNRAS.349.1153R}), where at low fluxes the reflection component is correlated with the power-law, while at higher fluxes the reflection component is more consistent with a constant. Also, we note that there appears to be a hysteresis behaviour in the relation at high fluxes (mainly in O4). If real (hard to say), it could be related to the geometry of emitting region.

In a reflection scenario, the reflected radiation is expected to respond to the primary power-law variations, and a positive correlation is expected. The fact that this is not the case at high fluxes is usually explained by light bending effects (\citealt{2003MNRAS.340L..28F,2004MNRAS.349.1435M}). The variability is driven mainly by a primary source moving in the strong gravity field. This can produce up to an order of magnitude flux variations with small changes in the reflected flux. The decreasing strength of the features at 1 and 7 \keV is similar to that seen in MCG-6-30-15 (\citealt{2003MNRAS.339.1237V}) and is in line with the predictions of the light bending model (\citealt{2004MNRAS.349.1435M}).

\begin{figure}
\centering
\includegraphics[width=200pt,clip ]{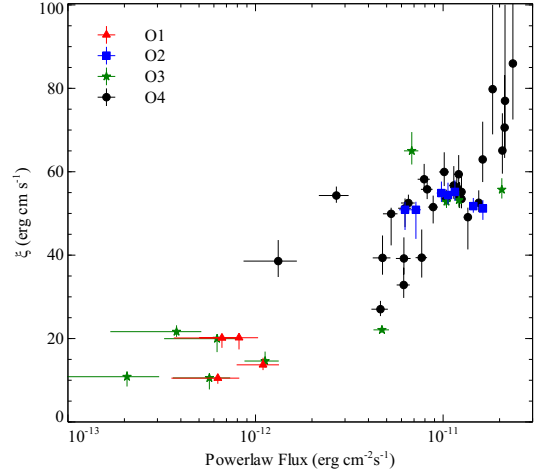}
\caption{Ionisation of the reflector vs power-law flux (0.3--10 \keV) in the 10 ks segments fits (see also Fig. \ref{fig:pow_ref_all} and text).}
\label{fig:pow_xi_all}
\end{figure}

The other interesting result is related to the ionisation changes. Fig. \ref{fig:pow_xi_all} shows the variations in ionisation parameter $\xi$ along with the power-law flux, and shows that the ionisation parameters changes as the power-law flux changes. In agreement with the interpretation of the difference spectrum of Fig. \ref{fig:diff_spec} (see Sec. \ref{discussion} for more interpreting this in the light of the light bending model). The ionisation changes of the reflector between 10--100 ($\rm{erg\ cm\ s}^{-2}$) causes the shape of the reflection spectrum (that is possibly otherwise constant) at $\sim 1$ keV to change, it is mainly caused by changes in the iron L emission, and also possible changes from other elements, and this is responsible for the `bump' in Fig. \ref{fig:diff_spec}.

\section{Timing Analysis}\label{timing}
To further understand the properties of the source, we focus in this section on the timing properties to establish whether or not it is consistent with our interpretation of the spectrum and spectral variability.

One key prediction of the reflection interpretation of the spectrum is a time delay (lag) between the primary illuminating power-law and the emission from the reflector acting as a mirror.
\subsection{Coherence}
The coherence of two light curves is a measure of how correlated they are, or how much of one light curve can be predicted from the other. Mathematically, it is defined as:
\begin{equation}
\gamma ^2(f) = \dfrac{|\langle C(f)\rangle|^2}{\langle|S(f)|^2\rangle\langle|H(f)|^2\rangle}
\end{equation}
where $S(f)$ and $H(f)$ denotes the Fourier transforms of the two light curves under study. $C(f) = \langle S^*(f)H(f)\rangle$ is the cross spectrum, and $^*$ denotes the complex conjugate. The angled brackets indicates the average over several light curves and/or frequency points.

The coherence takes a value of 1 if the two light curves are perfectly coherent (for details on the meaning of coherence and how it is calculated, see \citealt{1997ApJ...474L..43V, 1999ApJ...510..874N, 2003MNRAS.339.1237V}).

\begin{figure}
\centering
\includegraphics[width=240pt,clip ]{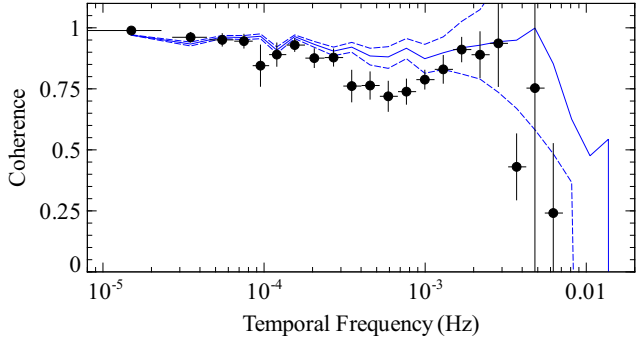}
\caption{Black dots: Coherence as a function of Fourier frequency (inverse of time-scale) between the two energy bands 0.3--1.0 (soft) and 1.0--4.0 \keV (hard). The blue line is the mean of the coherence of 1000 simulated light curve pairs that have the same properties as the observed data. The dashed lines represent the 95 per cent confidence levels. The frequency bins were constructed so that the bin size is equal to 1.4 times the frequency value}
\label{fig:coherence}
\end{figure}

Fig. \ref{fig:coherence} shows the coherence function $\gamma^2(f)$ between the two energy bands 0.3--1.0 (soft) and 1.0--4.0 \keV (hard). To calculate this, background-subtracted light curves were extracted in the two bands, then divided into 4 segments of about 100 ks length each. The Fourier transform of each segment was then taken and the coherence and its errors were calculated following equation 8 in \cite{1997ApJ...474L..43V}. The frequency bins were constructed so that the bin size is equal to 1.4 times the frequency value. The figure shows that the two light curves are highly coherent ($\sim 1$) for almost all of the frequency range of interest. The coherence drops sharply at $\sim 5\times 10^{-3}$ Hz where the noise starts to dominate.

To confirm this we simulated 1000 light curve pairs (each representing the two energy bands), that have the same statistical properties as the observed data (see Sec. \ref{simulations} on details about the simulations). We followed the method of \cite{1995A&A...300..707T} to generate red-noise light curves that have the same mean and variance as the data, then applied Poisson noise and the same windowing and binning functions as that used for the real data. The simulated light curves are perfectly coherent and the only source of deviation from 1 in the coherence function is due to Poisson noise. As can be seen is Fig. \ref{fig:coherence}, where the blue continuous line is the median of the coherence measured from 1000 light curves, the data is fully consistent with the simulated light curves, which imply that the light curves of interest are high coherent up to $\sim 5\times 10^{-3}$ Hz. 

\subsection{Time Lags}\label{time_lags}
\begin{figure}
\centering
\includegraphics[width=240pt,clip ]{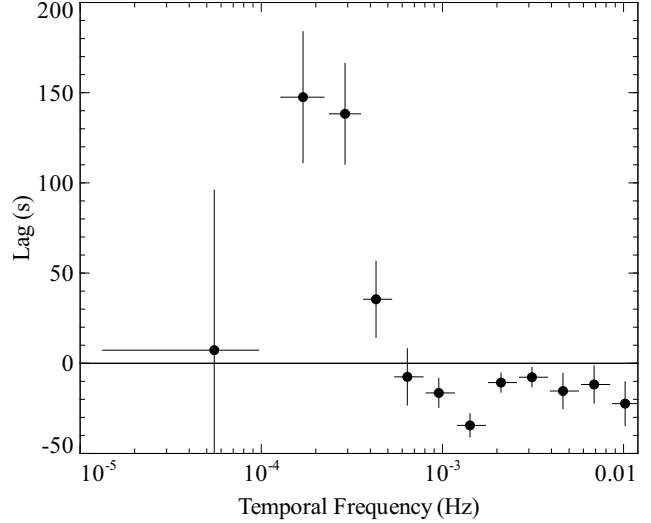}
\caption{Time lag as a function of Fourier frequency (inverse of time-scale) between the two energy bands 0.3--1.0 (soft) and 1.0--4.0 \keV (hard). The frequency binning is similar to that of Fig. \ref{fig:coherence}. A positive lag (mainly below $\sim 5\times10^{-4}$ Hz) indicates that the hard flux lags behind (i.e. changes after) the soft flux.}
\label{fig:time_lag}
\end{figure}

Fig. \ref{fig:time_lag} shows the time lag between the 0.3--1.0 and 1.0--4.0 \keV energy bands as a function of Fourier frequency (similar to Fig. 3 in \citealt{2009Natur.459..540F}). A similar procedure to the coherence was followed. The lag $\tau(f)$ was calculated following \cite{1999ApJ...510..874N} with

\begin{equation}
\tau(f) = \dfrac{\phi(f)}{2\pi f}=\dfrac{\rm{arg}[C(f)]}{2\pi f}
\end{equation}
where $C(f)$ is again the cross spectrum, and $\rm{arg}[C(f)]$ is the argument of the complex number $C(f)$.
The sign convention here means that a positive lag indicates the soft flux changes \textit{before} the hard flux. The errors in the lag were calculated again following equation 16 in \cite{1999ApJ...510..874N}, and are investigated using Monte Carlo simulations in Sec. \ref{simulations}.

The figure shows that below $\sim 5\times10^{-4}$ Hz, the soft flux leads the hard flux by about 150 seconds, the lag appears to turn over below $\sim 10^{-4}$ Hz. At $\sim 6\times10^{-4}$ Hz, the lag turns negative, and variations in the hard band now lead. However, above $\sim 5\times10^{-3}$ Hz, the coherence in the light curves is lost (Fig. \ref{fig:coherence}) and the lag is dominated by Poisson noise. The shape of the lag function cannot be characterised by a single power-law as it is commonly done with other sources (e.g \citealt{2001ApJ...554L.133P, 2003MNRAS.339.1237V,2006MNRAS.372..401A}). However, if fitted only to the range $[2\times10^{-4}-3\times10^{-3}]$ Hz, a power-law gives a good description with the best fit function being $\tau(f) = 0.34f^{-0.77}-87$.

The lag measurements could be affected by artefacts and biases in the Fourier calculations, and Poisson noise could also have a contribution. To investigate the significance of the lag measurements, the best procedure is to use Monte Carlo simulations, and that is the topic of the following section.

\subsection{Lag Significance}\label{simulations}
The idea here is to generate light curves with the same statistical properties as the data, and then impose a defined lag and see how well it can be recovered.

The light curves were generated from red-noise power spectra (PSD) following the method of \cite{1995A&A...300..707T}. The PSD of the data was calculated using light curves from the whole energy band, and was fitted with a broken power-law. The best fitting model had an index of $2.18\pm0.09$ at high frequencies breaking to $0.8\pm0.5$ at a frequency of $f_{\rm break}=1.4^{+0.9}_{-0.4}\times10^{-4}$ Hz (if the low frequency index is fixed at 1, the break is $1.6^{+0.4}_{-0.6}\times10^4$ Hz). This best fitting model was used as the underlying PSD in generating the simulated light curves. Using energy-resolved PSDs instead of the total PSD does not change the results of the simulation.

One thousand light curve pairs were generated in each experiment, where for each light curve pair, a random realisation of the Fourier transform for one light curve was generated. The second light curve Fourier transform was produced from this by adding a frequency-dependent phase to the Fourier transform. They are then inverse Fourier-transformed to yield two light curves that have the desired lag function between them. The light curves were scaled so they have the same mean and variance as that of the data before adding Poisson noise. To account for any bias in the data that usually are associated with Fourier transforms, mainly windowing and aliasing effects, the generated light curves were longer than the data, they were then resampled to match the data.
\begin{figure}
\centering
\includegraphics[width=240pt,clip ]{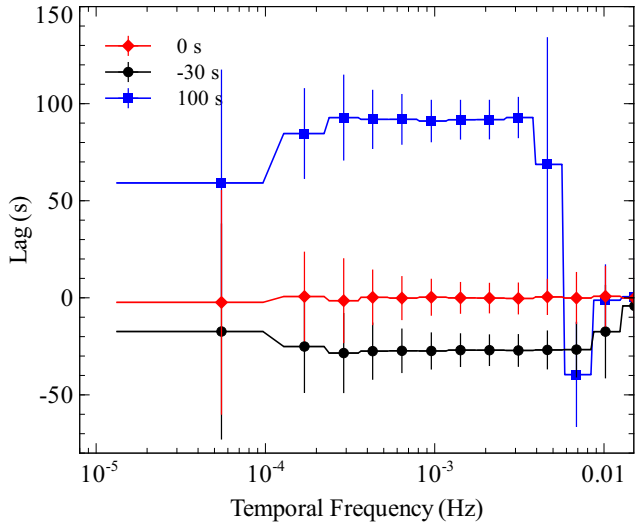}
\caption{Time lag as a function of Fourier frequency for three sets of simulated light curves. The time lag was calculated from each light curve and then averaged. The error bars are the standard deviation in the spread of the measured lags. Blue squares are for a lag of 100 s, red diamonds are for 0 s and black circles are for -30 s lag.}
\label{fig:sim_lag_3const}
\end{figure}

Fig. \ref{fig:sim_lag_3const} shows the result of averaging 1000 lags measured for three sets of light curves, with input lags of 0, $-30$ and $+100$ seconds, so they roughly cover the range of lags observed. The error bars represent the $1\sigma$ spread in the measured lags from different light curves. The lags can be recovered to a very good degree in the middle frequencies $5\times 10^{-4} - 5\times10^{-3}$ Hz, with the probability of measuring a lag that is off by more than $\sim 15s$ from the true value is less than 5 per cent.

At higher frequencies, as we saw in Fig. \ref{fig:coherence}, the light curves coherence drops below 1 sharply, and the noise dominates the data. For the 100 s lag, the sharp drop in the lag is caused by the fact that the lag is comparable to the frequency over which it is measured (i.e. the lag is of the same order as the time-scale it occurs). The drop happens when the phase $\phi$ suddenly jumps from $\pi$ to $-\pi$, which for 100 s lag occurs at $f=\phi/(2\pi\tau) = 5\times10^{-3}$ Hz. This is also the reason the lag tends to zero at high frequency. Phase flipping between $\pi$ and $-\pi$ makes the lag averages to zero. At lower frequencies, the spread in the measured lag (or equivalently, the errors in Fig. \ref{fig:time_lag}) becomes large, and that is because the time-scale probed now are comparable to the segments size over which the Fourier transform was performed. Also, it should be noted that spread in the lag measurements is fully consistent with the error formula in \cite{1999ApJ...510..874N} (their equation 17).

\begin{figure}
\centering
\includegraphics[width=160pt,clip ]{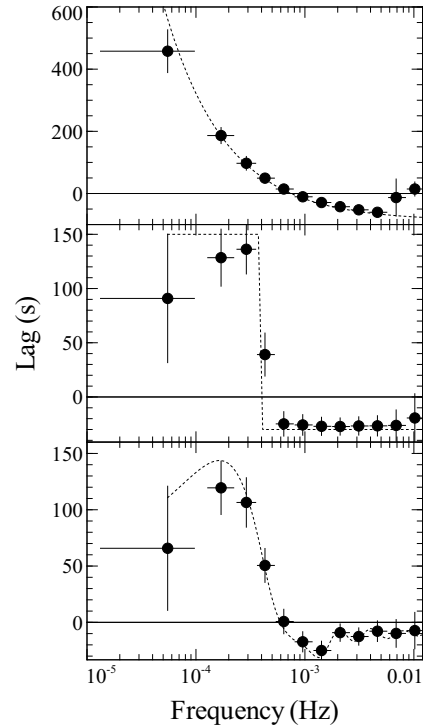}
\caption{Time lag as a function of Fourier frequency for three sets of simulated light curves. The input time lag functions are (from top to bottom): a power-law of the form $\tau(f) = 0.34f^{-0.77}-87$, a step function that changes from $+150 $ s to $-30$ s at a frequency of $4\times10^{-4}$ Hz, and an interpolated function describing the measured lag of Fig. \ref{fig:time_lag}. The points are the mean of the measured lag from 1000 light curve pairs. The error bars represent the one sigma spread. The input lag function is plotted as a dotted line in each case.}
\label{fig:sim_lag_3func}
\end{figure}
\begin{figure*}
\centering
\includegraphics[width=310pt,clip ]{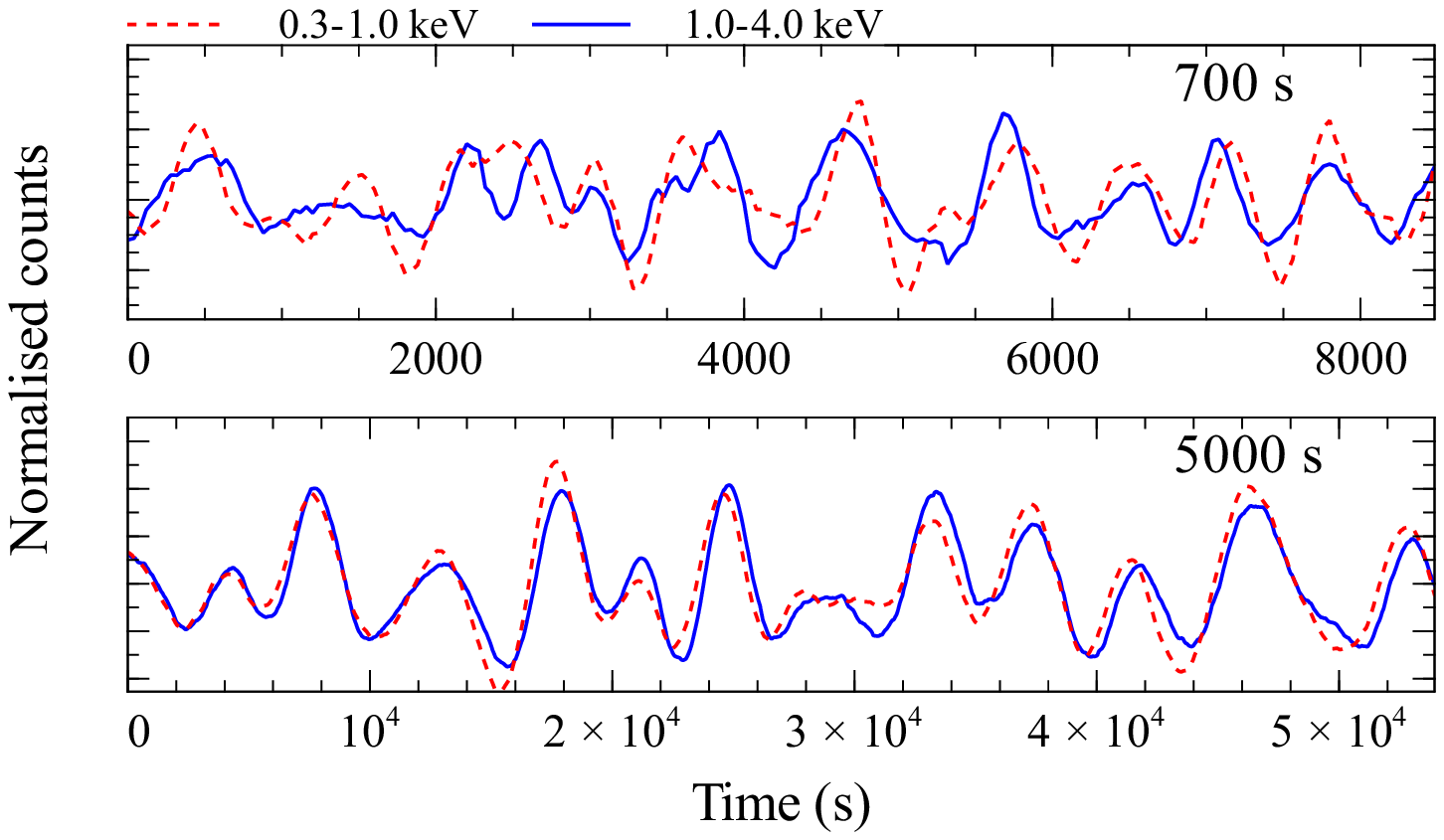}
\caption{Two segments of the light curves at 0.3--1.0 (red dotted line) and 1.0--4.0 \keV (blue continuous line). Top: variations on time-scale of 700s, corresponding to the lowest negative point in Fig. \ref{fig:time_lag}. Bottom: variations on time-scale of 5000s, corresponding to the highest positive point in Fig. \ref{fig:time_lag}. Notice the time-scale on the x-axis.}
\label{fig:show_lag}
\end{figure*}
\begin{figure}
\centering
\includegraphics[width=240pt,clip ]{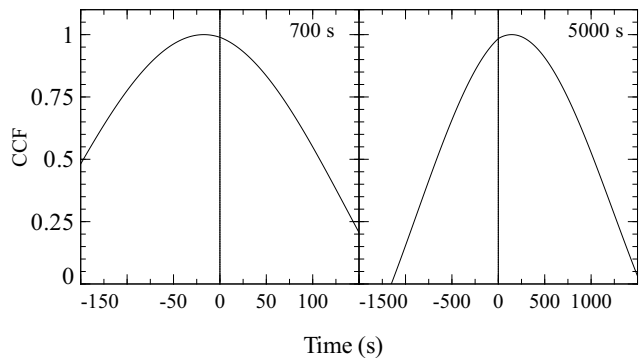}
\caption{The average CCF (see method (i) for details), for two time-scales, 700s and 5000s, corresponding to the minimum and maximum in Fig. \ref{fig:time_lag}. The vertical dotted line is the zero lag line. This clearly shows that the CCF is distorted towards negative lags for the 700 s case, and towards positive lags for the 5000 s time-scale. Notice the difference in x-axis scaling.}
\label{fig:show_corr}
\end{figure}

To see if the shape of the lag function itself (other than constant) has any effect on the measurements, we repeated the simulation for three lag functions: a power-law of the form $\tau(f) = 0.34f^{-0.77}-87$ which was found to fit the central frequencies of the measured lag, a step function that changes from $+150 $ s to $-30$ s at a frequency of $4\times10^{-4}$ Hz, and finally an interpolated smooth function describing the measured lag. The result are shown in Fig. \ref{fig:sim_lag_3func}. The lag functions are clearly well recovered, with the earlier mentioned trends visible at very high and low frequencies, with the lag at the very low end slightly under estimated.

The main conclusion is that for most of the frequency range of interest, the spread in the simulated light curve lags caused by the noise is small and the lags measured in Fig. \ref{fig:time_lag} are significant. In particular the soft lag which is the first to be seen in active galaxies (\citealt{2009Natur.459..540F}, see \citealt{2004MNRAS.347..269G} for a possible soft lag in IRAS 13224-3809).

\subsection{Alternative Approach: Time domain}
A significant lag in the data, in particular the soft lag, would in principle be visible in the time domain light curves. The problem however, is apparent from Fig. \ref{fig:time_lag}, and that is, a raw light curve would, if anything, show the average lag over \textit{all} time-scales. In order to get any useful lags in the time domain, the time-scale (or a frequency range) has to be selected first. This is achieved by smoothing the light curve by two gaussians of slightly different sizes, then subtracting them, so that only variations on the time-scale of interest are left. This is the 1-dimensional equivalent of unsharp masking in image processing. The shape of the smoothing function is unimportant for the study presented here, and several functions were tested and they do not alter the conclusions drawn\footnote{Some functions produce biases in the lag that can be calibrated using Monte Carlo simulations}.
\begin{figure}
\centering
\includegraphics[width=230pt ]{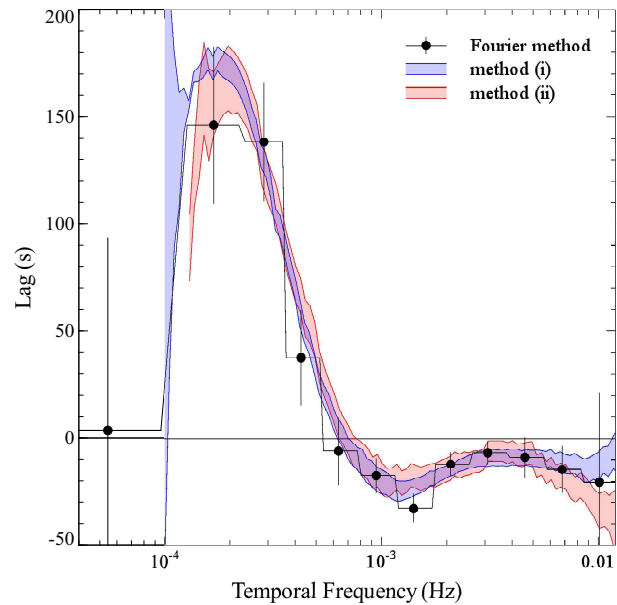}
\caption{The same as Fig. \ref{fig:time_lag}, with the result of the two time domain methods over-plotted. Black dots represent the lag measured using the phase difference in the Fourier domain. The shaded areas represent the $1\sigma$ error for the lag measured using the methods discussed in Sec. \ref{Tmethods}. The lag was calculated for 100 time-scales between $10^2$ and $10^4$ s, with the errors coming from simulations of 50 light curve pair for each point.}
\label{fig:show_lagT}
\end{figure}

Fig. \ref{fig:show_lag} shows two segments of the light curves at the energies of interest (0.3--1.0 and 1.0--4.0 \keV). In the top panel, the time-scale corresponds to the lowest negative point in Fig. \ref{fig:time_lag}, and it shows how the dotted line (0.3--1.0 \keV) lags behind the continuous curve (1.0--4.0 \keV), and this is the negative soft lag seen in Fig. \ref{fig:time_lag}. The trend is reversed in the bottom panel when a different time-scale is selected. Now the dotted line leads, and this corresponds to the positive lag at low temporal frequencies.

To quantify this lag in a systematic way, we employed two methods that measure the lag as a function of time-scale. These are essentially equivalent to measuring it through phase lags using Fourier transforms.
\begin{enumerate}\label{Tmethods}
\item \textit{Correlation:} After selecting the time-scale of interest (e.g. Fig. \ref{fig:show_lag}), the light curve is split into segments whose length is several times the time-scale (e.g. for a time-scale of 700 s, the length of the segments is 7000 s). The cross correlation function (CCF) between light curves in the two energy bands (0.3--1.0 and 1.0--4.0 \keV) is then calculated for each segments. The CCFs from the segments are averaged, and the lag is measured through the maximum of the average CCF (see for example Fig. \ref{fig:show_corr}).
\item \textit{`Flare' Stacking:} In this method, again after selecting the time-scale in the light curves of the two bands (0.3--1.0 and 1.0--4.0 \keV) using a smooth function (e.g. Fig. \ref{fig:show_lag}), all peaks in one of the two light curves (which is used as a reference) are identified. For each peak, the corresponding `\textit{flare}' is identified by selecting the points either side of the peak. These segments of the reference light curve are matched by their time equivalent from the second light curve (not necessarily representing flares).
The lag is found by comparing the average of all flares (.i.e an `average flare'), with the corresponding (average) segment from the second light curve (this is to say, on average, what does a flare in one band correspond to in the other band).
\end{enumerate}

In order to obtain an estimate of the errors of the lag measured using the two previous methods, we again used Monte Carlo simulations, where for each measured lag, 50 light curve pairs that have lag equivalent to the measurements are generated, the same procedure of Sec. \ref{simulations} is used, the error is taken as the $1\sigma$ spread in the distribution of the measured lags.\\
Fig. \ref{fig:show_lagT} shows the result of doing the calculations using the two methods for 100 time-scales between $10^2$ and $10^4$ seconds, the shaded areas represent the errors taken as the $1\sigma$ spread in the lag calculation from simulated 50 light curve pairs for each point. This method was tested extensively on simulated data with known lags, and was found to recover the lags very well. This figure shows a very good match between the methods and further emphasises the significance of the lag measurement.
\section{Discussion}\label{discussion}
A detailed analysis of a long XMM observation of 1H0707 has been presented. It is clear that the source during the present O4 observation is in a state similar to the earlier O2 reported in \cite{2004MNRAS.353.1064G}. The high energy spectral drop has an energy of $7.31^{+0.12}_{-0.06}$ \keV. The two models, partial covering and reflection, both give an equally good fit to the data above 3 \keV. However, as noted in \cite{2009MNRAS.tmpL.244R} for MCG-6-3-15, an absorber that is responsible for the spectral drop is expected to produce fluorescent emission at $\sim 6.4$ \keV. This is clearly not seen in the data, and might require a special geometry, where the absorber covers only a small part of the sky as seen by the source. The value of $\sim 20$ per cent covering fraction for the absorber at the source, is an upper limit, and although it is not as small as that reported in \cite{2009MNRAS.tmpL.244R}, it points to the fact that a `special geometry' explanation cannot hold for all AGN. Since, to reproduce the drop, most of the absorber must lie exactly in our line of sight with a covering fraction of 70 per cent, and a global covering fraction (i.e. as seen by the source) of about 20 per cent.

\begin{figure}
\centering
\includegraphics[width=180pt ]{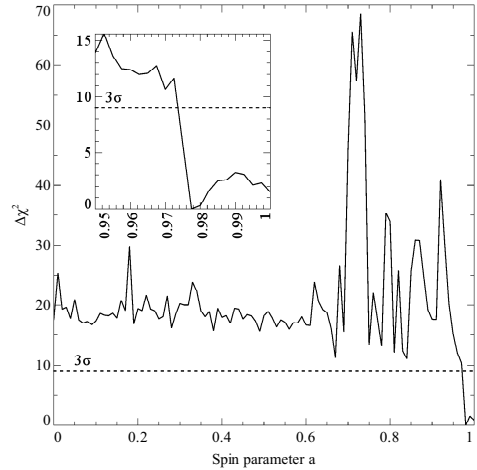}
\caption{$\Delta\chi^2$ versus spin parameter for a fit using \texttt{kerrdisc}. The black hole is rapidly spinning. The fit was made to 3.0-10.0 \keV band for the combined O4 observations. \textit{Inset:} a zoom-in on the high spin values. The dashed line in both plots marks the $3\sigma$ limit.}
\label{fig:spin}
\end{figure}

The drop is more consistent with being the blue wing of a broad K$\alpha$ emission line, originating very close to the black hole.
The inner radius of the emitting region is $\sim1.3 r_{\rm g}$, and if we assume no radiation is emitted from within the radius of marginal stability, the black hole has to be an almost maximally-spinning Kerr black hole. A fit using \texttt{kerrdisc} model from \cite{2006ApJ...652.1028B} gives a spin parameter $a>0.976$ at 90 per cent confidence (where a is the dimensionless parameter $a=cJ/GM^{2}$, with $J$ being the angular momentum of a black hole of mass $M$, see \citealt{2009ApJ...697..900M} for spin measurements using this method for stellar mass black holes). Fig. \ref{fig:spin} shows the $\chi^2$ changes for the spin parameter. The parameter is well constrained to be $>0.97$, and this is the case also if it is measured from the low flux observations (O1 and the first of O3). Although this might not be an exact value (\citealt{2006ApJ...652.1028B}, submitted), what is clear from the shape of the line is that most of the radiation has to be emitted from within a few gravitational radii, and the black hole in 1H0707 appears to be rapidly spinning similar to what was found for MCG-6-30-15 (\citealt{2006ApJ...652.1028B}).

This interpretation is further strengthened with the identification of the feature at $\sim 1$ \keV as the a broad iron L-shell line (\citealt{2009Natur.459..540F}). As we have shown in Sec. \ref{soft_fit} (see Fig. \ref{fig:base_plus_diff}), the spectral complexities around 1 \keV cannot be fitted with any absorption-only model, and there is a requirement for an emission component. The flux ratio between the K and L features is 1:20, measured as the ratio of equivalent widths of two \texttt{laor} lines, is consistent with the value expected from atomic physics (\citealt{2005MNRAS.358..211R}).

It is apparent from the models fitted to the high energy band that the spectrum requires high iron abundance. This is expected to imprint other signatures in the spectrum. In particular, if the apparent drop at $\sim 1$ \keV is interpreted as an iron L edge, then a deep absorption trough is expected to be produced by iron M-shell UTA. A reflection continuum on the other hand, accounts naturally for most of the features.

The high iron abundance ($\sim 9\times$ solar), if real, is not inconsistent with what is expected from line ratio modeling at other wavelengths (\citealt{2002ApJ...567L..19S,2002ApJ...564..592H}), and would imply that the host galaxy had gone through a phase of strong star formation. This seem to be consistent with studies showing that NLS1 have stronger star formation than normal Seyferts (\citealt{2009arXiv0908.0280S}). Also the FeII emission in also strong in these sources, and it can also be attributed to high abundance (\citealt{2000NewAR..44..531C}). The high abundance may also be due to a dense nuclear supernova activity. On the modeling side, non-solar values for elements other than iron can slightly affect the shape of the broad iron K$\alpha$ and the abundances inferred from the fits (\citealt{1995MNRAS.276.1311R}).

Because the source is highly variable, fitting a time-averaged spectrum might not give the whole picture, Both flux and time-resolved spectroscopy indicate that the spectrum is composed of two components, a highly variable component between 1--4 \keV and a less variable component outside the range. These two components are consistent with a PLC and RDC respectively, similar to the results from other AGN (\citealt{2003MNRAS.340L..28F,2003MNRAS.342L..31T,2004MNRAS.348.1415V}). In particular, the shape of the constant (or slowly varying) component dominating the low flux state matches very well the shape of a reflection spectrum (see Fig. \ref{fig:spec_all}). Variations in the power-law component drives most of of the variability (Fig. \ref{fig:pow_ref_all}), and this simple model reproduces the variability pattern very well (Fig. \ref{fig:flux_resolved_image}). 

No absorption-only model provides a good fit to the spectrum, and these models can also be ruled out based on spectral variability. If the spectral drop at $\sim 7$ \keV is due to an absorbing cloud in the line of sight, then the depth and shape of the drop imply the absorber only partially covers the source, which in turn means it has to be separate from any absorber causing the drop at $\sim 1$ \keV. This however, is inconsistent with the variability pattern showing that the spectral features at $\sim 7$ and $\sim 1$ \keV are varying in a similar way (Fig. \ref{fig:flux_resolved_image}).

The variability pattern, where a good correlation between reflection and the direct continuum is seen at low fluxes and levels off at higher fluxes, is consistent with the predictions of the light bending model (\citealt{2003MNRAS.340L..28F,2004MNRAS.349.1435M}). Most of the luminosity emerges from within few gravitational radii, and strong gravitational light bending \textit{must} be taken into account. However, the model in its simplest form assumes that the intrinsic luminosity of the source is constant, which most likely is not the case. Also, assuming a single ionisation parameter throughout the disc might not be realistic too. These two factors may be responsible for producing the positive correlation in Fig. \ref{fig:pow_xi_all}, where an increase in the intrinsic luminosity of the primary component causes the disc ionisation to increase. The most likely scenario is that both light bending and intrinsic source variations are at work.
\subsection{Time Lag and Reverberation}
The lag spectrum is frequency dependent, and cannot be fitted with a single power-law. If however, the fit is only applied to the central frequency bins, a good fit is found with $\tau(f)\propto f^{-0.77}$. The slope cannot be constrained very well, and a power-law with an index of 1 is also consistent with the data. This is similar to (or at least consistent with) what was found in NGC 7469 (\citealt{2001ApJ...554L.133P}), MCG-6-30-15 (\citealt{2003MNRAS.339.1237V}), NGC 4051 (\citealt{2004MNRAS.348..783M}) and NGC 3783 (\citealt{2005ApJ...635..180M}). Several models have been suggested to explain the lag, the most successful of which is the propagation of accretion fluctuations (\citealt{1997MNRAS.292..679L,2001MNRAS.327..799K}, see also \citealt{2006MNRAS.367..801A}). In this model, accretion rate variations are produced in different radii in the accretion disc, each with time-scale associated with the radius of origin. These fluctuations propagate inward and modulate the central X-ray emitting region. This seems to explain most of the timing properties including the PSD shape, its changes with energy and RMS-flux relation (\citealt{2004MNRAS.347L..61U}). If additionally, energy-dependent emissivity profiles are assumed, with soft energies having flatter emissivity profiles, the dependency of the lag on frequency and energy can also explain the observations in galactic black holes and AGN.

The lag is produced when large time-scale fluctuations propagating inward passes through the soft-emitting region first (larger radii), before reaching the hard-emitting region (\citealt{2001MNRAS.327..799K}). This produces the positive lag below $\sim 5\times10^{-4}$ Hz. The magnitude of the lag itself is dependent on the propagation speed and the distance between the soft- and hard-emitting regions.

The slope of the lag spectrum appears to change at small temporal frequencies (large time-scales). If this turnover is real (i.e. the last point in Fig. \ref{fig:time_lag}), it might be the effect of the size of the emitting region. The emissivity profiles inferred from the spectral fitting to the iron line indicate that most of the radiation originates within few gravitational radii. This might pose an upper limit on the hard lags that can be observed. Interestingly, the maximum of the lag, at $1-2\times10^{-4}$ Hz, is about the same as the break frequency in the PSD, which was found to be $1.6^{+0.4}_{-0.6}\times10^{-4}$ Hz, (when fitted with a broken power-law). The break in the PSD in this propagation model may be related to the time-scale of the fluctuations at the outer radius of the X-ray emitting region (\citealt{2001MNRAS.327..799K,2004MNRAS.348..783M}). The light crossing distance corresponding to $\sim150$ s for a black hole mass of $5\times10^{6}\rm{M}_\odot$ is $\sim 8\ r_{\rm g}$. The viscous time-scale at this radius in a standard disc (\citealt{1973A&A....24..337S}) is $t_{\rm{visc}}\sim 10^{6}$ s (assuming a viscosity parameter $\alpha=0.1$ and a scale height $H/R =0.1$), which is much higher than what is observed. If however, thermal time-scales are considered where $t_{\rm{th}}=(H/R)^{2}t_{\rm{visc}}$ (which might be the case if variability is caused by ionisation changes), then the lag peak and the PSD break would correspond to the thermal time-scale of the outer edge of an emitting region that is less than $8 r_{\rm g}$ in size.

At time-scales below about half an hour, the lag becomes negative, where hard energies now \textit{lead} softer energies. This is the first significant soft lag detected from black holes (\citealt{2009Natur.459..540F}). Other authors have reported soft lags before, but they were not significant (e.g. \citealt{2007MNRAS.382..985M} for Ark 564 and \citealt{2004MNRAS.347..269G} for IRAS 13224-3809). If the positive lag is interpreted as being due to Comptonisation, where the variability is seen first in the soft band (disc component in the form of reflection) before the hard band (Comptonised coronal emission), then negative soft lag above $\sim 6\times10^{-4}$ Hz is simply in the wrong direction. More sophisticated models may be tuned to produce soft lags. In particular, \cite{2000A&A...359..843M} studied the temporal evolution of flares in a disc-corona system. They found that, if a flare originates in the corona (e.g. by magnetic reconnection) with the perturbation time-scale of the order of few corona light crossing time, then a soft lag is expected, and is caused by a feedback mechanism that cools the corona down. Fluctuations are seen in harder energies, before reprocessed soft photons from the disc cool the corona and softens the spectrum. In this picture, the whole spectrum originates in the corona. This however, is inconsistent with the spectral variability patterns seen in Sec. \ref{spec_var}. There is an abrupt change in variability properties between the bands above and below $\sim 1$ \keV (see also RMS spectrum in Fig. 8 of the supplementary material in \citealt{2009Natur.459..540F}), which argues for two \textit{separate} components above and below 1 keV.

In a reflection scenario, the soft lag is naturally expected. The variability is seen first in the power-law dominated component, emitted from the corona. Some of the radiation is reflected off the optically thick disc. This produces a lag comparable to the light travel time between the the corona and the reflector. If we assume most of the radiation is emitted at $\sim 2 r_{\rm g}$, implied from the inner radius and emissivity index in the spectral fitting, then interpreting the 30 s lag as the light crossing time yields a black hole mass of $\rm{M}\sim2\times10^6 \rm{M}_\odot$, which is consistent with the uncertain mass of this black hole quoted in the literature (e.g. \citealt{2005ApJ...618L..83Z}).

Ark 564 shows a similar lag spectrum to the one presented here (\citealt{2007MNRAS.382..985M}), with a difference in the magnitude of the lag. The ratio of hard to soft lags in Ark 564 is $\sim700/10=70$, compared to $\sim5$ in 1H0707. This appears to be related to the compactness of the emission region and the strength of the reflection. 1H0707 has a strong reflection component associated with a compact emission region with strong light-bending effects. However Ark 564 has weaker reflection which suggests a more extended emitting region.  Since thermal and viscous time-scales scale as $(r/r_g)^{3/2}$, the hard lags due to changes in the accretion flow in this region are larger compared to the light-travel time.

Unlike other well-studied AGN (e.g. MCG-6-30-15 and Mrk 766), the spectrum of 1H0707 does not show any signs of warm absorption, giving an uninterrupted view to the inner regions of the accretion system. 1H0707 has one of the strongest soft excesses among other NLS1 (e.g. \citealt{2006MNRAS.368..479G}). That, in our interpretation, is simply because of the high iron abundance making the iron L complex strong, which might also be linked to the conclusion of \cite{2006MNRAS.368..479G}, that most NLS1 with complex spectral features show high values of Fe~\textsc{ii}/H$\beta$. Other sources that are similar to 1H0707 include IRAS 13224-3809, which shows clearly the iron L broad feature, however the current XMM observation is not long enough to check if there is any soft lags (Ponti et al. 2009, submitted).

In summary, although different models may be made to explain some of the properties we see independently, most of them fail to explain all of what we see. In this work, we have explored the data using different methods, and shown that a model in which the contribution of X-ray reflection off optically thick material is significant, provides a natural explanation to the spectrum, spectral variability, as well as the soft lag seen for the first time in 1H0707.
\section*{Acknowledgements}
AZ thanks the Cambridge Overseas Trust and STFC. ACF thanks the Royal Society for support. PU acknowledges support from an STFC Advanced Fellowship. GP thanks ANR for support (ANR-06-JCJC-0047). GM thanks the Spanish Ministerio de Ciencia e Innovaci\'{o}n and CSIC for support through a Rom\'{o}n y Cajal contract. CSR thanks the US National Science Foundation for support under grant AST06-07428.

\bibliography{bibliography}
\end{document}